\documentclass{iau}
\usepackage{graphicx}

\title[Optical selection of quasars] 
{Optical selection of quasars: SDSS and LSST}

\author[\v{Z}eljko Ivezi\'{c} et al.]   
{\v{Z}eljko Ivezi\'{c}$^1$,
W. Niel Brandt$^2$,
Xiaohui Fan$^3$,
Chelsea L. MacLeod$^4$,
Gordon T. Richards$^5$,
\and Peter Yoachim$^1$
}

\affiliation{
$^1$ Department of Astronomy, University of Washington, \\ Box 351580, Seattle, WA 98195-1580, USA\\ 
                               email: {\tt ivezic@astro.washington.edu} \\[\affilskip]
$^2$  Department of Astronomy and Astrophysics, The Pennsylvania State University, \\ 525 Davey Laboratory, University Park, PA 16802, USA\\
                               email: {\tt niel@astro.psu.edu} \\[\affilskip]
$^3$  Steward Observatory, University of Arizona, \\ 933 North Cherry Avenue, Tucson, AZ 85721, USA\\
                               email: {\tt fan@as.arizona.edu} \\[\affilskip]
$^4$  Department of Physics, U. S. Naval Academy, \\ 022 Chauvenet Hall, Annapolis, MD 21402, USA\\ 
                               email: {\tt macleod@usna.edu} \\[\affilskip]
$^5$  Department of Physics, Drexel University, \\ 3141 Chestnut Street, Philadelphia, PA 19104, USA\\ 
                               email: {\tt gtr@physics.drexel.edu}
}

\pubyear{2014}
\volume{304}  
\pagerange{1--7}
\jname{Multiwavelength AGN Surveys and Studies}
\editors{A. Mickaelian, F. Aharonian \& D. Sanders, eds.}

\begin{document}
\maketitle

\begin{abstract}
Over the last decade, quasar sample sizes have increased from several thousand to several hundred 
thousand, thanks mostly to SDSS imaging and spectroscopic surveys. LSST, the next-generation 
optical imaging survey, will provide hundreds of detections per object for a sample of more than 
ten million quasars with redshifts of up to about seven. We briefly review optical quasar selection techniques, 
with emphasis on methods based on colors, variability properties and astrometric behavior.
\end{abstract}

\firstsection 

\section{Introduction}

The selection of large samples of active galactic nuclei, including quasars as their high-luminosity tail, 
is required in many astrophysical areas, such as galaxy evolution, black hole growth, and the large-scale structure 
of the universe. The available quasar samples have increased by over two orders of magnitude in less than two decades
and this rapid progress is expected to continue (for example, simulations predict that LSST 
will deliver a sample of about 10 million quasars; see Section~\ref{sec:lsst}). Here we briefly review optical quasar 
selection methods based on SDSS data and discuss how they will be improved with the aid of more precise 
time-resolved photometry expected from LSST.

\section{Finding quasars with SDSS}

The similarity of unresolved quasars to stars in optical imaging surveys poses a difficulty in their identification. 
Quasars can be reliably identified by their spectra, and the SDSS provided a spectroscopically complete survey to 
$i < 19$ with $\sim$106,000 quasars (Schneider et al. 2010). This is the largest homogeneous quasar sample with 
high-quality optical spectra assembled to date (P\^{a}ris et al. 2012 recently reported an additional $\sim$74,000 
objects selected with selection criteria modified to enable the Lyman-$\alpha$ forest analysis).
These objects were selected as quasar candidates using colors measured from the $ugriz$ SDSS imaging data, and 
also using radio continuum 20 cm data from the FIRST survey (Becker et al. 1995). After SDSS collected enough 
multi-epoch imaging data, it was demonstrated that photometric variability-based selection is even more efficient 
than color-based selection.  These two selection methods are briefly summarized here.

\subsection{Color selection of quasar candidates}

\begin{figure}[t]
\vskip -2.0in
\begin{center}
 \includegraphics[width=0.99\textwidth]{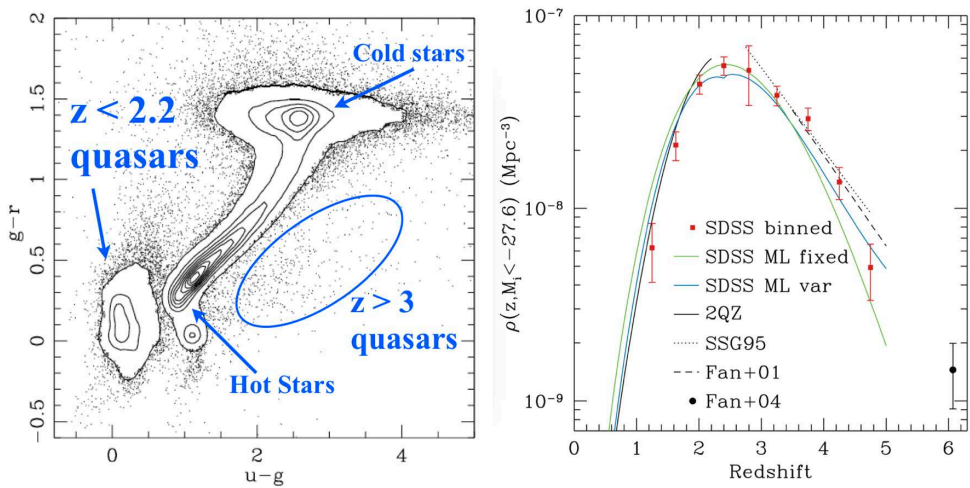} 
\vskip -2.2in
 \caption{
The left panel illustrates the color-based selection method for quasar candidates. Essentially,
the algorithm selects all sources whose colors place them outside the main stellar locus, 
seen in the middle of the panel. Quasars with redshifts below 2.2 have distinctively blue $u-g$
colors. The right panel shows the variation of volume number density of luminous quasars. 
The local maximum at about $z=2.4$ is clearly visible. Adapted from Richards et al. (2006).}
\label{fig:UVX}
\end{center}
\end{figure}

A small fraction of SDSS sources unresolved in imaging data were automatically targeted as quasar candidates for 
spectroscopic followup (Richards et al. 2002). The quasar targeting algorithm selects all point sources with $15 < i < 19.1$ 
($i$ is the apparent  magnitude in the SDSS $i$ band) whose colors place them outside the main stellar locus (it is a bit more 
complicated than this -- for details please see figure 1 in Richards et al. 2002). For quasars at redshifts below about two, 
the most discriminatory color is the $u-g$ color: compared to stars of the same visual color (e.g., the $g-r$ color), quasars 
show an excess of ultraviolet flux in the $u$ band and thus have bluer $u-g$ colors (see the left panel in Figure~\ref{fig:UVX});
this method is often called the UV-excess selection. The colors of quasars at higher redshifts are typically significantly 
different from stellar colors. A thorough analysis of the colors of quasars in the SDSS photometric system was presented in Richards et al. 
(2001). In addition, all unresolved sources from the same magnitude range that are within 2 arcsec of a FIRST radio 
detection are also targeted (for analysis of these radio quasars, see Kimball et al. 2011 and references therein). Some 
quasars were also targeted fortuitously via the algorithms for selecting galaxies (Strauss et al. 2002). 
The completeness of the resulting quasar sample is above 90\% (the confirmed fraction of all quasars within the adopted 
flux limits and within the surveyed area) and the efficiency
of color selection is about 65\% (that is, about 35\% of selected quasar candidates turned out not to be quasars). 
This homogeneously selected sample spans a large redshift range (there are 56 quasars at redshifts beyond 5 in
the Schneider et al. sample) and has enabled numerous quasar studies. For example, the peak in the comoving 
volume number  density of luminous quasars (dominated by type I objects) is now reliably and precisely determined 
(see the right panel in Figure~\ref{fig:UVX}), and the luminosity functions of quasars and AGN galaxies (selected using 
emission line strengths) appear mutually consistent despite grossly different selection procedures (Hao et al. 2005). 
We note that several hundred candidate type II quasars (high-luminosity analogs of type 2 Seyfert galaxies) were
found in the SDSS spectroscopic survey of galaxies (Zakamska et al. 2003, 2004).  

The relatively simplistic color-based selection algorithm employed by the original SDSS spectroscopic target
selection pipeline has been significantly improved using modern data mining and machine learning methods.
For example, Richards et al. (2009) introduced a kernel density estimator and a non-parametric Bayesian 
classification method, and Bovy et al. (2012) introduced a Gaussian mixture model to recognize quasar
candidates. These methods have yielded large samples of candidates (of order a million) and with improved
completeness and efficiency tradeoff; Richards et al. (2009) reported efficiency of up to 97\% while maintaining
fairly high completeness levels above 70\%. More details about the performance of these modified selection
algorithms are available in P\^{a}ris et al. (2012) and Ross et al. (2012).

\begin{figure}[t]
\vskip -2.0in
\begin{center}
 \includegraphics[width=0.99\textwidth]{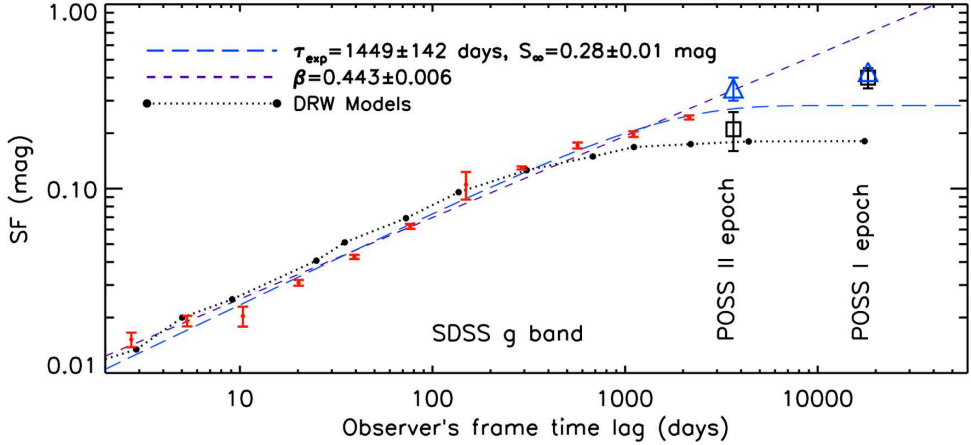} 
\vskip -2.3in
 \caption{Structure function for quasar variability measured in the SDSS $g$ band and in the observer's frame. 
The small dots with error bars represent the SDSS measurements, and the data points at $\Delta t > 3000$ days 
(large squares and triangles) are inferred from comparing SDSS data and POSS I/II data. The short-dashed line 
shows the best power-law fit to the SDSS measurements alone, and the long-dashed line shows a simultaneous
fit to all data points using the structure function expected for a damped random walk (see text). The best-fit parameters 
for these two fits are listed in the upper-left corner. The dotted line shows the prediction of the damped random walk 
model trained using light curves for quasars from Stripe 82. Adapted from MacLeod et al. (2012).}
\label{fig:SFinf}
\end{center}
\end{figure}

\begin{figure}[t]
\vskip -0.8in
\centering
  \begin{tabular}{@{}cc@{}}
  \hskip -0.4in
    \includegraphics[width=.59\textwidth]{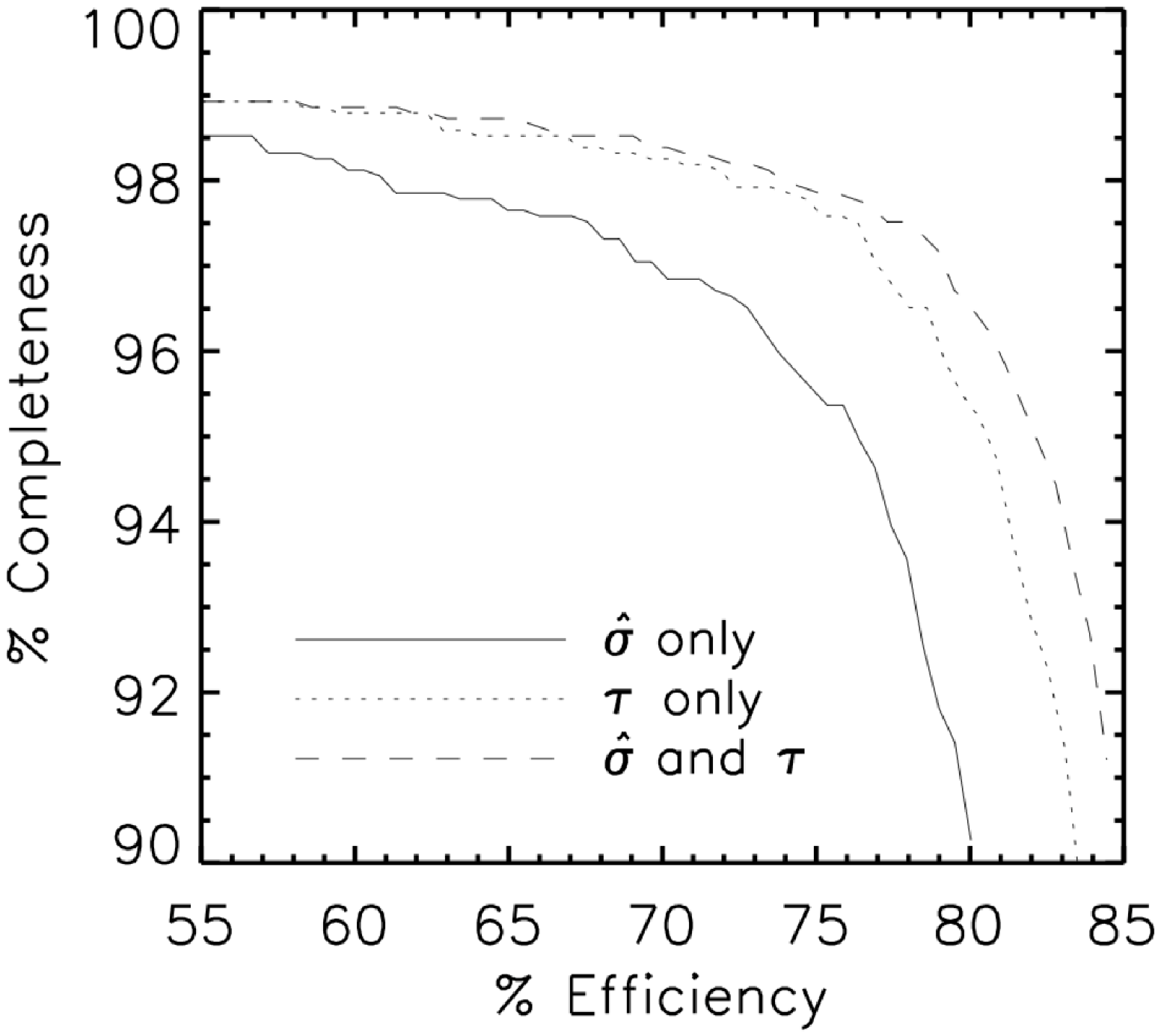} &
  \hskip -0.58in
    \includegraphics[width=.59\textwidth]{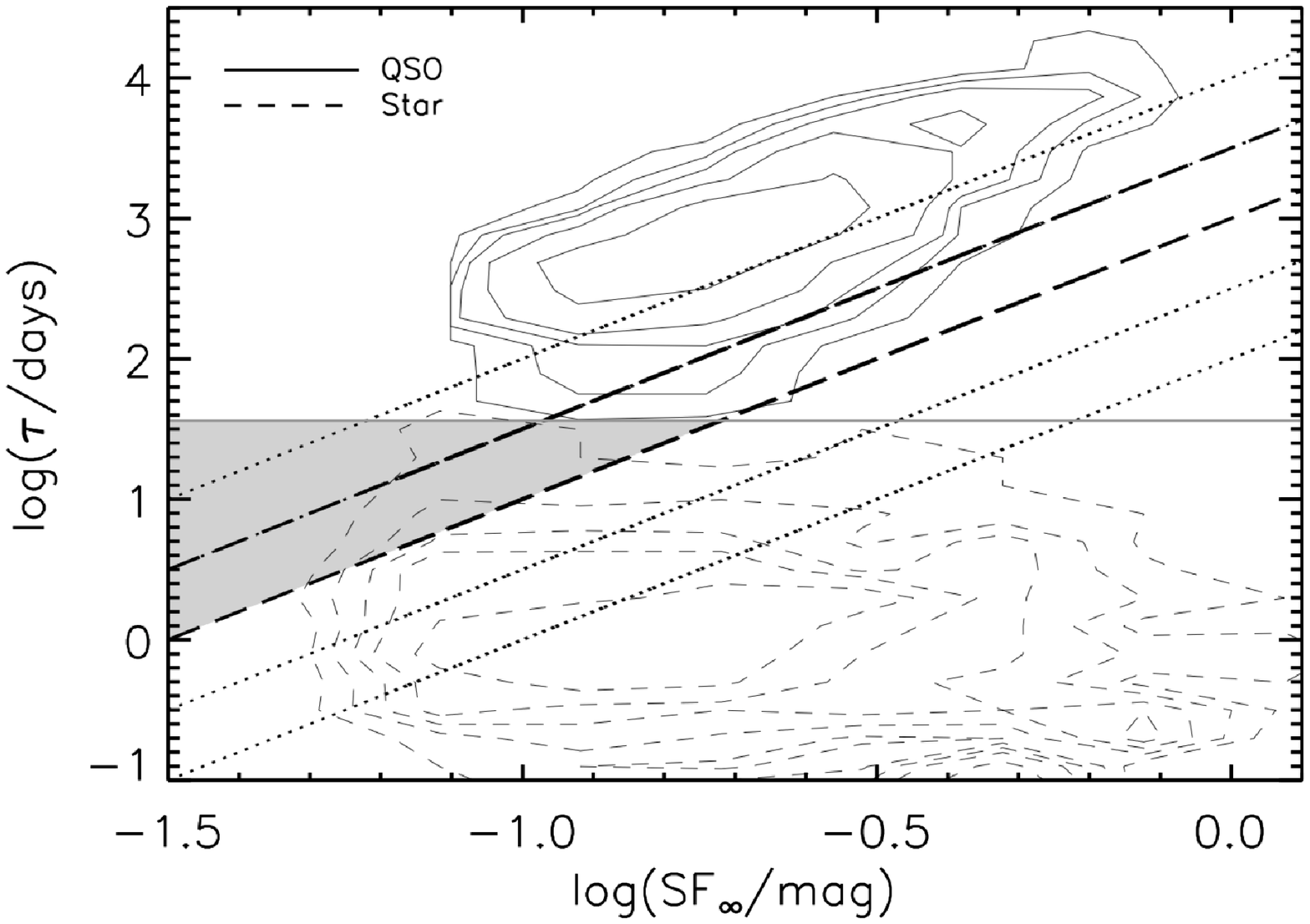} \\
  \end{tabular}
\vskip -1.0in
\caption{Illustration of improvements in variability-based selection due to added time-scale information. 
The solid line in the left panel shows the maximum completeness as a function of efficiency, for quasars 
from SDSS Stripe 82 region with $i<19$, achieved when using only the amplitude of 
the structure function for short time lags ($\hat{\sigma}= {\rm SF}_\infty / \sqrt{\tau}$). The dashed
line shows how efficiency improves when also including a cut in characteristic time scale ($\tau$). 
The dotted line shows results when using the $\tau$ information alone. The right panel shows the 
distribution of spectroscopically confirmed quasars (solid contours, enclosing 40\%, 60\%, 75\%, 85\%, 
and 90\% of data points) and stars (dashed contours) in the time scale ($\tau$) vs. asymptotic variability 
(SF$_\infty$) diagram for objects from SDSS Stripe 82 with $i<19$.  The dotted lines represent 
lines of constant $\hat{\sigma}$. The two thick dashed lines correspond to $\hat{\sigma} = 10^{-0.22}$ 
and $10^{-0.47}$ mag yr$^{-1/2}$. The gray region represents the contaminating (stellar) region when 
selecting sources with $\hat{\sigma} < 10^{-0.22}$ mag yr$^{-1/2}$. When imposing a lower limit at 
$\tau = 10^{1.56} = 36.3$ days (gray horizontal line), these contaminants are excluded from the sample, 
leading to a higher efficiency of quasar selection. Adapted from MacLeod et al. (2011).}
\label{fig:MacLeod}
\end{figure}

\subsection{Variability selection of quasars}

Quasars are variable sources with optical amplitudes of several tenths of a magnitude for time scales longer than 
a few months, and this behavior can be used for their selection (Hawkins \& Veron 1995; Ivezi\'{c} et al. 2003). 
Sesar et al. (2007) showed using SDSS Stripe 82 data (a $\sim$300 deg$^2$ equatorial region imaged about 60 times) 
that practically all quasars spectroscopically confirmed by SDSS are also variable in SDSS imaging data, and 
Koz{\l}owski et al. (2010) demonstrated that variability can be used to separate quasars from most variable stars 
even in the dense stellar environments of the Magellanic Clouds.

A number of studies used SDSS light curves for close to 10,000 quasars from Stripe 82 to quantify the structure 
function for quasar variability (Ivezi\'{c} et al. 2004; MacLeod et al. 2010; Schmidt et al. 2010; Butler \& Bloom 2011; 
Palanque-Delabrouille et al. 2011). The structure function as a function of time lag $\Delta t$, SF($\Delta t$),
as defined in recent quasar studies, is equal to the standard deviation of the distribution of the magnitude difference 
$m(t_2)-m(t_1)$ evaluated at many different times $t_1$ and $t_2$, such that time lag $\Delta t = t_2-t_1$
(and divided by $\sqrt{2}$ because of differencing). The structure function is directly related to the autocorrelation 
function, which makes a Fourier pair with the power spectral density function (PSD). When the structure function 
${\rm SF} \propto (\Delta t)^\alpha$, then ${\rm PSD} \propto 1/f^{(1+2\alpha)}$.

The power-law index $\alpha$ is a good discriminator between variable stars and quasars\footnote{This
parameter is also a good discriminator of various models for the origin of quasar variability, see
Kawaguchi et al. (1998).} (Schmidt et al. 2010; Butler \& Bloom 2011). The key insight is that for time lags
below a year or so quasars have a much steeper 
structure function ($\alpha \sim 0.4-0.5$) than most variable stars. In other words, compared to their 
variability at a time lag of, say, one year, quasars vary by 1-2 orders of magnitude less at time lags
below a month or so -- this is not true for the overwhelming majority of variable stars. For UV-excess selected 
objects, variability-based methods that utilize $\alpha$ select quasars with a completeness of 90\% and a purity of 
95\% (Schmidt et al. 2010). Furthermore, this performance level is maintained in the redshift range $2.5 < z < 3$ 
where color selection is known to be problematic. 

The power-law dependence of the structure function for quasar variability cannot be extrapolated beyond
a few years, as shown using SDSS and POSS data by Ivezi\'{c} et al. (2003) and later using also Stripe 82 data by
MacLeod et al. (2012; see Figure~\ref{fig:SFinf}). It has been demonstrated that quasar variability can be 
statistically described using a stochastic model called damped random walk (Kelly, Bechtold \& Siemiginowska 
2009; Koz{\l}owski et al. 2010, MacLeod et al. 2010, 2011, 2012; Zu et al. 2012). Also known in the physics 
literature as the Ornstein--Uhlenbeck process, and as the continuous autoregressive process in the statistics 
literature, this model includes three parameters: the mean value (of the quasar magnitude), the characteristic 
time scale $\tau$, and the asymptotic (at time scales longer than $\tau$) root-mean-square variability,
SF$_{\infty}$.  Alternatively, the model (as well as its modifications) can be described via a covariance matrix 
(see the contribution in these Proceedings by Ivezi\'{c} \& MacLeod). The predicted structure function for the 
damped random walk process is ${\rm SF}(\Delta t) = {\rm SF}_\infty \, \left[1 - \exp(- \Delta t / \tau) \right]^{1/2}$. 
At small time lags, ${\rm SF}(\Delta t) \propto \Delta t ^ {1/2}$, and thus a damped random walk is equivalent to 
an ordinary random walk for $\Delta t < \tau$ (for a random walk, ${\rm PSD} \propto 1/f^2$; the ``damped'' aspect 
manifests itself as a flat PSD for $\Delta t > \tau$). 

The time span of SDSS data from Stripe 82 is sufficiently long to constrain $\tau$ for the majority of the 
$\sim$10,000 quasars with light curves. MacLeod et al. (2011) showed that this additional time scale
information can be used to improve significantly the selection based only on the slope of the structure 
function for small time lags (see Figure~\ref{fig:MacLeod}). For example, it is possible to construct samples 
with only 15\% stellar contamination even when using {\it only} variability-based constraints (see the left 
panel in Figure~\ref{fig:MacLeod}), and 2\% stellar contamination when a UV-excess constraint is added
(see Figure 12 in MacLeod et al. 2011), while maintaining a completeness of 90\%.

\section{Finding quasars with LSST}

The last decade has seen fascinating observational progress in optical imaging surveys.
The SDSS dataset is currently being greatly extended by the ongoing surveys such as 
Pan-STARRS (Kaiser et al. 2010) and the Dark Energy Survey (Flaugher 2008). The Large Synoptic 
Survey Telescope (LSST; for a brief overview see Ivezi\'{c} et al. 2008) is the most ambitious 
survey currently planned in the visible band. LSST will extend the faint limit of SDSS by about 
5 magnitudes and will have unique survey capability in the faint time domain. In particular,
LSST will revolutionize our understanding of the growth of supermassive black holes with 
cosmic time, AGN fueling mechanisms, the detailed physics of accretion disks, the contribution 
of AGN feedback to galaxy evolution, the cosmic dark ages, and gravitational lensing
(for a detailed discussion, see Chapter 10 in the LSST Science Book, Abell et al. 2009). 

While no massive spectroscopic followup of quasar candidates will be attempted as
part of the LSST project, the time-resolved aspect of LSST photometric and astrometric data 
will enable significant improvements in the completeness and efficiency of resulting quasar 
samples compared to single-epoch measurements. We first briefly describe anticipated LSST 
surveys and then discuss how these data will be used to construct quasar samples with 
up to about 10 million objects.

\subsection{Brief overview of anticipated LSST surveys \label{sec:lsst}}.

The LSST design is driven by four main science themes: probing dark energy and dark matter, 
taking an inventory of the Solar System, exploring the transient optical sky, and mapping the 
Milky Way. LSST will be a large, wide-field ground-based system designed to obtain multiple 
images covering the sky that is visible from Cerro Pach\'{o}n in Northern Chile. The project is 
scheduled to have first light around 2019 and the beginning of survey operations by 2021. 

The current baseline design, with an 8.4m (6.5m effective) primary mirror, a 9.6 deg$^2$ field 
of view, and a 3.2 Gigapixel camera, will allow about 10,000 deg$^2$ of sky to be covered 
using pairs of 15-second exposures in two photometric bands every three nights on average, 
with typical 5$\sigma$ depth for point sources of $r\sim24.5$. The system is designed to 
yield high image quality as well as superb astrometric and photometric accuracy. The survey 
area will include 30,000 deg$^2$ with $\delta<+34.5^\circ$, and will be imaged multiple times 
in six bands, $ugrizy$, covering the wavelength range 320--1050 nm. About 90\% of the 
observing time will be devoted to a deep-wide-fast survey mode which will observe an
18,000 deg$^2$ region over 800 times (summed over all six bands) during the anticipated 
10 years of operations, and yield a coadded map to $r\sim27.5$. These data will result in 
databases including about 20 billion galaxies and a similar number of stars, and will 
serve the majority of science programs. The remaining 10\% of the observing time 
will be allocated to special programs such as a Very Deep and Fast time-domain 
survey. More details about various science programs that will be enabled by LSST data
can be found in the LSST Science Book (Abell et al. 2009).

\subsection{Color and variability selection of quasars}

The existing selection algorithms based on colors and photometric variability developed with the aid of SDSS data 
will be applicable to LSST data, too. Although LSST will not obtain simultaneous multi-band photometry
like SDSS did, the averaging of many observations (ranging from about 50-60 in the $u$ band to 
180-190 in the $r$ and $i$ bands) will result in sufficiently precise color measurements to 
recognize easily  color offsets from the main stellar locus. The addition of variability information will 
boost the sample efficiency to levels comparable to those obtained for spectroscopic samples. 
Two additional selection methods, enabled by the multi-epoch LSST imaging and described
below, will further improve the resulting samples. 

Detailed simulations of the quasar luminosity function and light curves, the LSST observing cadence, and 
the LSST photometric error distribution (Abell et al. 2009; MacLeod et al. 2011; Palanque-Delabrouille 
et al. 2013) show that the LSST quasar sample will include close to 10 million objects, and will be complete 
for $M < -23$ objects (a formal absolute magnitude definition cutoff of quasars) to redshifts beyond 3. 
Notably, LSST will discover about 1000 quasars with redshifts beyond 7 (using the $z$-band
drop-out technique) which will represent a valuable sample for studying the epoch of reionization.

\subsection{Additional constraints: proper motion and differential chromatic refraction} 

In addition to photometry, astrometric measurements can help to distinguish quasars from 
stars. First, measurable proper motion will reject about 2/3 of all stars, even before any color or 
photometric variability criteria are applied. LSST proper motion errors will be 0.5 mas/yr for 
sources with $r=23$ and 1.0 mas/yr for $r=24$ (Ivezi\'c et al. 2008).  Simulations of the proper
motion distribution for Milky Way stars (for model description, see Ivezi\'{c}, Beers \& Juri\'{c} 2012)
indicate that over $\sim$80\% of stars with $r<23$ and over $\sim$70\% of stars with $r<24$ 
will have proper motions three times larger than expected measurement errors (with very little 
dependence on Galactic coordinates). These high fractions of rejecteable stars are also expected for 
various stellar subpopulations, such as brown dwarfs (contaminants for very high-redshift quasar candidates) 
and white dwarfs (contaminants of $z<2.2$ candidate samples). 

An additional astrometric effect that is sensitive to detailed differences in the spectral energy
distributions between quasars and stars is the differential chromatic refraction (DCR). Due to
the wavelength dependence of atmospheric refraction, in images astrometrically calibrated
using stars, quasars show astrometric offsets in a given bandpass as a function of airmass, 
and also show astrometric offsets between different bandpasses even at a fixed airmass (if larger
than one, of course). As shown in Figure~5 from Kaczmarczik et al. (2009), the astrometric offsets 
of quasars from Stripe 82 between the $u$ and $g$ bands, and between the $g$ and $r$ bands, 
can be up to about 20 mas (depending on redshift), even for moderate airmass ($\sim$1.1-1.2). 
The measurement errors for these offsets anticipated for the LSST main survey will be below 3 mas
for objects brighter than $r=22$, about 5 mas for objects with $r\sim23$, and 10-15 mas for 
objects with $r\sim24$. Therefore, the DCR offsets for quasars will represent an additional 
quantity, indpendent of color, variability and proper motion, to help distinguish quasars from stars.

Last but not least, both color and DCR measurements can be used to 
estimate quasar redshifts with a precision comparable to that expected for LSST galaxies. 
Using SDSS data, Kaczmarczik et al. (2009) obtained a measurement precision for this 
``photo-astro-z'' of 0.03, with the fraction of quasars with redshifts correct to within 0.1
of order 90\%.

\end{document}